\documentclass[aps,prd,onecolumn,eqsecnum,amsmath,nofootinbib,preprintnumbers]{revtex4}%
\newcounter{example}[section]

\usepackage{appendix}
\usepackage{amsmath}
\usepackage{color,graphicx,float,subfigure}
\usepackage{amsfonts,amssymb,theorem,mathrsfs,times}
\usepackage{bm}
\usepackage{ulem}
\usepackage{amsfonts,amssymb,theorem,mathrsfs}
{\theorembodyfont{\upshape}
	}
{\theorembodyfont{\upshape}
	}
{\theorembodyfont{\upshape}
	}
{\theorembodyfont{\upshape}
	}
{\theorembodyfont{\upshape}
	}
{\theorembodyfont{\upshape}
	}

\newcommand{\dalm}{\kern1pt\vbox{\hrule height 0.9pt\hbox{\vrule width
			0.9pt\hskip 2.5pt\vbox{\vskip 5.5pt}\hskip 3pt\vrule width
			0.3pt}\hrule height 0.3pt}\kern1pt}

\begin{document}
	\title{Robust topological invariants of timelike circular orbits for spinning test particles in black hole spacetimes}
	%
	
	\author{ Yong Song\footnote{e-mail
			address: syong@cdut.edu.cn (corresponding author)}}
	\author{Jiaqi Fu\footnote{e-mail
			address: 1491713073@qq.com}}
	\author{Yiting Cen\footnote{e-mail
			address: 2199882193@qq.com}}

	
	\affiliation{
	College of Physics\\
	Chengdu University of Technology, Chengdu, Sichuan 610059,
	China}
	

	\date{\today}
	
	\begin{abstract}
	The spin-curvature coupling in the Mathisson–Papapetrou–Dixon (MPD) formalism induces non-geodesic motion, shifting the orbital parameters of spinning test particles in black hole spacetimes. We investigate whether these quantitative shifts alter the qualitative, global structure of the orbit manifold. Using a topological approach, we study timelike circular orbits (TCOs) for spinning particles in static, spherically symmetric spacetimes. By constructing an auxiliary vector field, we compute the topological winding number $W$ in horizon-bounded regions of asymptotically flat, anti–de Sitter (AdS), and de Sitter (dS) backgrounds. We find that $W$ is robust against both the magnitude and direction of the particle’s spin: between two horizons, $W = -1$, guaranteeing at least one unstable TCO; outside the outermost horizon in asymptotically flat and AdS spacetimes, $W = 0$, enforcing that TCOs must appear in stable–unstable pairs or be absent. This spin independence reveals that the fundamental orbital structure is a property of spacetime geometry itself, not of the particle's spin. We validate this with quantitative examples in Schwarzschild, Schwarzschild-AdS, and Schwarzschild-dS spacetimes, showing explicit spin-induced TCO shifts while confirming the invariant topology. This result provides a topological foundation for interpreting gravitational waveforms from extreme mass-ratio inspirals involving spinning secondaries.
	\end{abstract}
	

	\maketitle


\section{Introduction}
Circular orbits in black hole spacetimes are essential tools for probing strong-field gravity, underpinning key astrophysical phenomena such as accretion disks, black-hole shadows, and gravitational-wave emission from extreme mass-ratio inspirals (EMRIs). Their study provides deep insight into the structure of spacetime and the behavior of matter in extreme environments.

In recent years, topological methods have offered a powerful, model-independent framework for analyzing the existence and stability of circular orbits. These approaches provide a geometric, global perspective on orbital structures that transcends the details of specific metrics or matter content. Cunha, Berti, and Herdeiro~\cite{Cunha:2017qtt} pioneered a topological framework to study light rings (LRs) or photon spheres (PSs) of ultra-compact objects, demonstrating that non-degenerate LRs always appear in pairs. They then proceeded to investigate the topological properties of the LRs outside a stationary, axisymmetric black hole~\cite{Cunha:2020azh}, proving that at least one unstable LR exists outside the black hole horizon for each rotation sense. Subsequent work, particularly by Wei et al.~\cite{Wei:2020rbh}, applied Duan’s topological-current $\phi$-mapping theory~\cite{Duan:1979ucg,Duan:1984ws} to classify circular orbits in asymptotically flat, AdS, and dS black holes, uncovering universal topological invariants associated with photon spheres. Duan's topological current $\phi$-mapping can also provide new insights into the study of black hole thermodynamics~\cite{Wei:2021vdx,Wei:2022dzw}.

Although significant progress has been made in understanding null circular orbits, the case of timelike circular orbits (TCOs) is more intricate, owing to their dependence on orbital parameters such as angular momentum. Earlier studies established that, for neutral particles with fixed angular momentum, TCOs generically arise in stable–unstable pairs~\cite{Wei:2022mzv,Ye:2023gmk}; this pairing is topologically enforced and independent of black-hole parameters. Building on this, several further studies have been carried out~\cite{Shahzad:2024oxb,Ye:2024sus,Song:2025vhw}, including the case of charged particles. It has been shown that the electric charge of the particle indeed alters the topological properties of its circular orbits~\cite{Ye:2024sus,Song:2025vhw}.

However, this framework is incomplete for describing realistic astrophysical systems. In EMRIs targeted by future gravitational wave detectors like LISA, the secondary object could be a spinning neutron star or black hole. Similarly, accretion disk dynamics and the motion of dark matter candidates with non-zero spin involve particles where spin-curvature coupling becomes significant. The motion of spinning test particles is governed by the Mathisson–Papapetrou–Dixon (MPD) equations~\cite{Mathisson:1937zz,Papapetrou:1951pa,Tulczyjew:1959,Taub:1964,Madore:1969,Dixon:1964,Dixon:1965:1,Dixon:1970:1,Dixon:1970:2,Dixon:1973:1,Dixon:1974:1,Dixon:1979,Obukhov:2015eqa,Semerak:1999qc,Hackmann:2014tga,Steinhoff:2009tk}, which extend geodesic motion to include dipole interactions with spacetime curvature. This coupling introduces quantitative shifts in orbital energies and radii~\cite{Jefremov:2015gza,Toshmatov:2019bda,Toshmatov:2020wky}, but its effect on the global topological structure of the orbit space remains unknown.

This gap in understanding motivates our central research question: does the particle's spin, like its charge, alter the topological properties of circular orbits? Although the behavior of effective potentials for spinning particles and the existence of circular orbits have been studied since the pioneering work of Tod, de Felice, and Calvani~\cite{Tod:1976ud}, and spin is known to produce quantitative shifts in orbital parameters~\cite{Toshmatov:2019bda,Toshmatov:2020wky}, it remains unknown whether it changes the fundamental topological invariants that characterize the global orbital structure. If these invariants are robust against spin, it would imply a deeper level of universality in orbital dynamics, dictated purely by spacetime geometry.

Understanding whether the topological structure of orbits is robust against spin is crucial for gravitational waveform modeling. If the topology depends on spin, then waveform templates must account for spin-induced topological transitions. Conversely, if it is spin-independent, then certain qualitative features of the inspiral are universal, simplifying template construction for detectors like LISA.

We address this gap by employing a topological approach to study the circular orbits of spinning test particles in static, spherically symmetric spacetimes. The central objective of this work is to determine whether the topological winding number $W$, which classifies the existence and stability of TCOs, remains invariant under the introduction of particle spin. Using the Tulczyjew spin-supplementary condition~\cite{Tulczyjew:1959}, we construct an auxiliary vector field whose zeros correspond to circular orbits. By computing $W$ in various regions of black-hole spacetimes, we demonstrate that the topological structure is indeed independent of both the spin magnitude and orientation. This represents a significant generalization of previous topological results and reveals a new layer of robustness in orbital dynamics.

The topological approach employed here offers distinct advantages over traditional potential analysis, particularly for spinning particles where the effective potential becomes increasingly complex and parameter-dependent. While previous studies have analyzed specific orbital configurations through numerical scanning of potential extrema, our method provides a global, parameter-independent classification. By constructing an auxiliary vector field whose zeros correspond to circular orbits, we extract topological invariants that characterize the existence and stability of orbits without requiring explicit solutions to the equations of motion. This approach reveals universal patterns that persist despite the nonlinearities introduced by spin-curvature coupling.

Our results show that the topological number $W$ is independent of both the magnitude and the orientation of the spin—that is, whether the particle is co-rotating or counter-rotating. Specifically, between two neighbouring horizons we obtain $W = -1$, indicating the presence of at least one unstable TCO. Outside the outermost horizon of asymptotically flat and AdS spacetimes we find $W = 0$, implying that any TCOs must appear in stable–unstable pairs or be absent altogether. These conclusions are corroborated by explicit examples in Schwarzschild, Schwarzschild-AdS, and Schwarzschild-dS spacetimes.

This study extends the topological classification of circular orbits to spinning test particles, revealing a robust topological structure that persists even in the presence of spin-induced non-geodesic effects. The invariance of topological winding numbers under spin perturbations represents a significant finding: although spin quantitatively shifts orbital radii and energies, it does not alter the fundamental topological structure of the orbit space. This robustness has important implications for gravitational waveform modeling, suggesting that certain topological features of inspiral trajectories may be more universal than previously appreciated.

The present paper is organized as follows: In Sec.~\ref{section2}, we give a brief review of the equations of motion of spinning extended test bodies, i.e., MPD equations, and introduce Wei et al.'s topological approach to the study of circular orbits. In Sec.~\ref{section3}, we study the topology of circular orbits for spinning test particles between two neighboring horizons. In Sec.~\ref{section4}, we study the topology of circular orbits for spinning test particles in asymptotically flat, AdS, and dS black holes. In Sec.~\ref{section5}, we summarize and discuss our results.  In this work, we use geometrized units, setting $G = c = 1$.


\section{Mathisson-Papapetrou-Dixon equations and topological approach for circular orbits}\label{section2}
In this section, we present the general formalism for a spinning test particle within the Mathisson-Papapetrou-Dixon (MPD) approximation, up to the pole-dipole order~\cite{Mathisson:1937zz,Papapetrou:1951pa,Dixon:1964,Dixon:1965:1,Dixon:1970:1,Dixon:1970:2,Dixon:1973:1,Dixon:1974:1,Dixon:1979} and review the topological approach for studying the properties of circular orbits. We consider a spinning particle moving in a static, spherically symmetric spacetime, with the line element:
\begin{eqnarray}
	\label{metric}
	ds^2=-f(r)dt^2+f(r)^{-1}dr^2+r^2(d\theta^2+\sin^2\theta d\phi^2)\;,
\end{eqnarray}
where $f$ is a function of the radial coordinate $r$. The equations of motion for spinning test particles up to the pole-dipole order are given by the MPD equations:
\begin{eqnarray}
	\label{MPD1}
	&&\frac{DP^a}{d\lambda}=-\frac{1}{2}R^a{}_{bcd}u^b S^{cd}\;,\\
	\label{MPD2}
	&&\frac{DS^{ab}}{d\lambda}=2P^{[a}u^{b]}\;.
\end{eqnarray}
where $D/d\lambda$ is the covariant derivative along the particle’s trajectory with the affine parameter $\lambda$ given by $D/d\lambda \equiv u^a \nabla_a$, and $R^a{}_{bcd}$ is the Riemann tensor. The dynamical 4-momentum and kinematical 4-velocity of the particle are denoted by $P^a$ and $u^a$, respectively, and the anti-symmetric spin tensor is denoted by $S^{ab}$ (with $S^{ab} = -S^{ba}$).

To close the system of Eqs.~(\ref{MPD1}) and (\ref{MPD2}), a supplementary condition must be imposed. In this work, to restrict the spin tensor to generate rotations only, we use the Tulczyjew spin-supplementary condition~\cite{Tulczyjew:1959}, i.e.,
\begin{eqnarray}
	\label{TC}
	S^{ab} P_{b}=0\;.
\end{eqnarray}
From Eq.~(\ref{TC}), it follows that the canonical momentum and the spin of the particle provide two independent conserved quantities:
\begin{eqnarray}
	\label{ppm2}
	&&P^a P_{a}=-\mathcal{M}^2\;,\\
	\label{SSs2}
	&&\frac{1}{2}S^{ab}S_{ab}=S^2\;,
\end{eqnarray}
where $\mathcal{M}$ is the ‘dynamical’, ‘total’ or ‘effective’ rest mass of the body and $S$ is the spin length of the particle. The spin four-vector can be defined as
\begin{eqnarray}
	S^a=\frac{1}{2\mathcal{M}}\epsilon^{ba}{}_{cd}P_b S^{cd}\;,
\end{eqnarray}
where $\epsilon^{ba}{}_{cd}$ is the Levi–Civita tensor. It is easy to find out $S^a$ is orthogonal to $P^a$, i.e., $S^aP_a=0$. 

The Tulczyjew spin-supplementary condition (\ref{TC}) implies that the components of the 4-velocity $u^a$ are determined from the following relation~\cite{Kuenzle:1972uk}
\begin{align}
	\label{ua}
	u^a=\frac{m}{\mathcal{M}^2}\bigg(P^a+\frac{2S^{ab}R_{bcde}P^cS^{de}}{4\mathcal{M}^2+R_{abcd}S^{ab}S^{cd}}\bigg)\;,
\end{align}
where $m$ is a scalar parameter (the `kinematical' or `monopole' rest mass of a particle), and it is given by $u^aP_a=-m$. 

The conserved quantities associated with the spacetime symmetries via the Killing vectors $\xi^a$ can be expressed as
\begin{eqnarray}
	\label{conserved}
	P^a\xi_a-\frac{1}{2}S^{ab}\nabla_b\xi_{a}=P^a\xi_a-\frac{1}{2}S^{ab}\partial_b\xi_{a}=\mathrm{constant}\;.
\end{eqnarray}
Due to the spherical symmetry of the metric (\ref{metric}), and provided the spin vector is aligned perpendicular to the equatorial plane (parallel to the total angular momentum)~\cite{Hackmann:2014tga}, we can confine the motion to the equatorial plane, i.e., $\theta=\pi/2$. Along the worldline, there are two conserved quantities for the pole-dipole particle: the energy $E$ and the angular momentum $L$. From Eq.~(\ref{conserved}), the conserved quantities can be expressed as 
\begin{eqnarray}
	\label{E}
	&&-E=P_t+\frac{1}{2}f^\prime S^{tr}\;,\\
	\label{L}
	&&L=P_\phi+rS^{r\phi}\;.
\end{eqnarray}
where a prime denotes the derivative with respect to radial coordinate $r$. From the Tulczyjew spin supplementary condition (\ref{TC}), and noting that for motion confined to the equatorial plane the components, i.e., \(S^{\mu\theta}=0\)~\cite{Hackmann:2014tga}, we obtain:
\begin{eqnarray}
\label{Stphi}
&&S^{t\phi}=-\frac{P_r}{P_\phi}S^{tr}\;,\\
\label{Srphi}
&&S^{r\phi}=\frac{P_t}{P_\phi}S^{tr}\;.
\end{eqnarray}
 From Eq.~(\ref{ppm2}), we can get
\begin{eqnarray}
	\label{pr}
	(P^r)^2=P_t^2-f\bigg(\frac{P_\phi^2}{r^2}+\mathcal{M}^2\bigg)\;.
\end{eqnarray}
From Eq.~(\ref{SSs2}), and considering Eqs.~(\ref{Stphi}), (\ref{Srphi}) and (\ref{pr}),  one finds~\cite{Hackmann:2014tga,Toshmatov:2020wky}
\begin{eqnarray}
S^{tr}=\frac{sP_\phi}{r}\;,
\end{eqnarray}
where $s=S/\mathcal{M}$ is specific spin parameter. It should be noted that $s$ can have both negative and positive values depending on the direction of spin with respect to the direction of the orbital angular momentum $L$. While $L$ is conserved, $P_\phi$ is not necessarily constant, see Eq.~(\ref{L}). The sign of $s$ relative to $L$ defines co/counter-rotation. In the spherically symmetric case, $L$ can always be chosen to be greater than zero, i.e., $L > 0$. Then, when $s<0$, the spin and angular momentum are counter-rotating; when $s>0$, the spin and angular momentum are  co-rotating~\cite{Toshmatov:2020wky}. From the conservation of energy (\ref{E}) and angular momentum (\ref{L}), we have
\begin{eqnarray}
\label{pt}
&&P_t=-\frac{2rE+f'Ls}{2r-f's^2}\;,\\
\label{pphi}
&&P_\phi=\frac{2r(L+Es)}{2r-f's^2}\;.
\end{eqnarray}
To ensure the mathematical consistency of the derived expressions (e.g., avoiding divergences in $p_\phi$, see Eq.~(\ref{pt})) and to maintain the timelike character of the particle’s four-velocity $u^a$, see Eq.~(\ref{ua}), (thus preserving causal structure outside the event horizon), we restrict our analysis to cases where the spin magnitude is small enough such that:
\begin{align}
	\label{conditions}
	2r-f's^2>0\;.
\end{align}
This constraint is physically reasonable for test particles with spin magnitudes small compared to the black hole mass, which is the regime where the pole-dipole approximation remains valid. Putting eq.~(\ref{pt}) and (\ref{pphi}) into eq.(\ref{pr}), we get the result that
\begin{eqnarray}
\label{pr2}
	(P^r)^2\equiv\mathcal{V}=A(E-V_+)(E-V_-)\;,
\end{eqnarray}
where
\begin{eqnarray}
A=\frac{4(r^2-fs^2)}{(2r-f^\prime s^2)^2}\;,
\end{eqnarray}
and
\begin{eqnarray}
\label{Vpm}
V_{\pm}=\frac{(2f-rf^\prime)sL}{2(r^2-fs^2)}\pm\frac{2r-f^\prime s^2}{2(r^2-fs^2)}\sqrt{f[L^2+\mathcal{M}^2(r^2-fs^2)]}\;.
\end{eqnarray}
Let $B=r^2-fs^2$. At the horizon, where $f=0$, we have $B>0$. Outside the horizon, based on the assumption (\ref{conditions}), we have $B'=2r-f^\prime s^2>0$. Therefore, $B$ is monotonically increasing outside the horizon, which implies that $B>0$ everywhere outside the horizon. Consequently, $A>0$ everywhere outside the horizon. For $(P^r)^2\ge 0$, the energy must satisfy:
\begin{eqnarray}
E\ge V_+\;,\quad \mathrm{or}\quad E\le V_-\;,
\end{eqnarray}
We focus on the case of positive energy, which corresponds to the effective potential $V=V_+$. The conditions for circular orbits, $\mathcal{V}=\partial_r\mathcal{V}=0$, become:
\begin{align}
E=V\;,\quad\mathrm{and}\quad \partial_rV=0\;.
\end{align}
Similar to Ref.~\cite{Wei:2022mzv}, $\partial_r V$ depends on $r$, $L$, and $s$, {\color{blue}}a regular potential function, which is independent of the orbital parameters, cannot be introduced here, as discussed in~\cite{Cunha:2017qtt,Cunha:2020azh}. For simplicity, we do not consider the four-dimensional expressions involving $\theta$. Instead, following the approach of choosing auxiliary functions in Ref.~\cite{Wei:2021vdx}, we introduce a new function
\begin{align}
\label{Vtheta}
V_{\theta}=\frac{1}{\sin\theta}V\;.
\end{align}
where the factor $\frac{1}{\sin\theta}$ is an auxiliary term\footnote{The choice of the factor $\frac{1}{\sin\theta}$ is not unique; any function that maintains a universal asymptotic behavior at $\theta=0,\pi$ and preserves the correspondence between zeros of $\phi^a$ and equatorial circular orbits would yield the same topological conclusions. For instance, a factor like $1/\sin^2\theta$ or other powers of 
$\sin\theta$ could also be used, provided the resulting vector field remains well-behaved at the boundaries.}. This construction is motivated by the need to define a vector field $\phi^a$ (see below) that is regular on the entire $(r,\theta)$ plane, including the poles $\theta=0,\pi$, and whose zeros correspond to circular orbits on the equatorial plane. The factor $1/\sin\theta$ ensures the asymptotic behavior of $\phi^a$ at the boundaries $\theta=0,\pi$ is universal and independent of $L$ and $s$, which is crucial for a consistent topological classification.

In order to give a global topology, we require that the values of the angular momentum and spin do not change the asymptotic behavior of $\partial_r V_\theta$ at the boundary of the $(r, \theta)$ plane. Following the Ref. \cite{Wei:2020rbh}, we deﬁne a vector field $\phi=(\phi^r,\phi^\theta)$
\begin{align}
	\label{phi}
	\phi^r=\frac{\partial_r V_\theta}{\sqrt{g_{rr}}},\quad\phi^\theta=\frac{\partial_\theta V_\theta}{\sqrt{g_{\theta\theta}}}\;,
\end{align}
in a ﬂat vector space. From Eqs.~(\ref{Vpm}), (\ref{Vtheta}) and (\ref{phi}), we have
\begin{align}
	\label{phir0}
	\phi^r&=\frac{1}{4(r^2-s^2f)^2}\bigg\{2Ls\sqrt{f}[(2f-rf')(f's^2-2r)+(r^2-s^2f)(f'-rf'')]\nonumber\\
	&-2f[(2r-s^2f')^2+(r^2-s^2f)(s^2f''-2)]\sqrt{L^2+\mathcal{M}^2(r^2-s^2f)}\nonumber\\
	&+\frac{(r^2-s^2f)(2r-s^2f')[(L^2+r^2\mathcal{M}^2)f'+2\mathcal{M}^2f(r-s^2f')]}{\sqrt{L^2+\mathcal{M}^2(r^2-s^2f)}}\bigg\}\;,
\end{align}
and
\begin{align}
\label{phit0}
\phi^\theta=-\frac{\cos\theta}{r\sin^2\theta}\bigg[\frac{(2f-rf^\prime)sL}{2(r^2-fs^2)}+\frac{2r-f^\prime s^2}{2(r^2-fs^2)}\sqrt{f(L^2+\mathcal{M}^2(r^2-fs^2))}\bigg]\;.
\end{align}
It follows that 
\begin{align}
\partial^\mu V_\theta\partial_\mu V_\theta=(\phi^r)^2+(\phi^\theta)^2=||\phi||^2\;,
\end{align}
where $||\phi||=\sqrt{\phi^a\phi_a}$. So, $\phi^a=0\Leftrightarrow ||\phi||=0$. Thus, $\phi^a=0$ corresponds to the circular orbits of spinning test particles and $\theta=\pi/2$ as expected. Deﬁne an angle $\Omega$ such that 
\begin{align}
\label{angle}
\phi^r=\phi\cos\Omega\;,\quad \phi^\theta=\phi\sin\Omega\;,\quad \Omega=\arctan(\phi^\theta/\phi^r)\;.
\end{align}
For each circular orbit, we assign a topological charge. Following Refs. \cite{Duan:1979ucg,Duan:1984ws,Wei:2020rbh}, the topological current is:
\begin{align}
\label{jmu}
	j^\mu=\frac{1}{2\pi}\epsilon^{\mu\nu\rho}\epsilon_{ab}\partial_\nu n^a\partial_\rho n^b\;,
\end{align}
where $x^\mu=(t,r,\theta)$ and the unit vector $n^a=(n^r,n^\theta)=(\phi^r/||\phi||,\phi^\theta/||\phi||)$. This current is conserved and can be written as:
\begin{align}
	j^\mu=\delta^2(\phi)J^\mu\bigg(\frac{\phi}{x}\bigg)
\end{align}
with the Jacobi tensor:
\begin{align}
\epsilon^{ab}J^\mu\bigg(\frac{\phi}{x}\bigg)=\epsilon^{\mu\nu\rho}\partial_\nu\phi^a\partial_\rho\phi^b\;.
\end{align}
Note that $J^\mu$ is nonzero only at the zeros of $\phi$.  The topological current density is:
\begin{align}
	j^0=\sum_{i}^{N}=\beta_i\eta_i\delta^2(\vec{x}-\vec{z}_i)\;,
\end{align}
where the Hopf index ($\beta_i$) and the Brouwer degree ($\eta_i$) of the $i$-th zero point are expressed. The topological number is:
\begin{align}
\label{topological number1}
	W=\int_{\Sigma}j^0d^2x=\sum_{i}^{N}\beta_i\eta_i=\sum_{i}^{N}w_i\;,
\end{align}
where $w_i$  is the winding number of the $i$-th zero point within region $\Sigma$. 

\begin{figure}[H]
	\centering
	\includegraphics[width=3in]{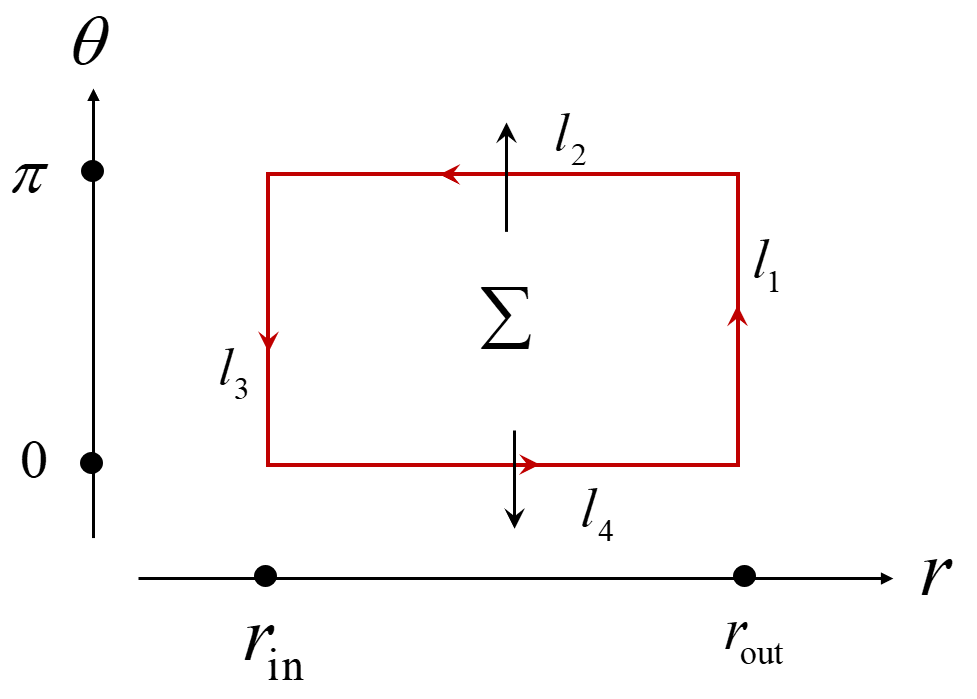}
	\caption{Representation of the contour $C=\sum_i\cup l_i$ (which encloses $\Sigma$) on the $(r,\theta)$ plane. The curve $C$ has a positive orientation, marked with the red arrows. $r_{\mathrm{in}}$ and $r_{\mathrm{out}}$ have different meanings in different cases. The black arrows indicate the approximate directions of the vector $\phi$ at the boundaries. At $\theta=0$ and $\pi$,  the direction of the vector $\phi$ is vertically upward and downward, respectively.}
	\label{C}
\end{figure}

The topological number $W$ defined in Eq.~(\ref{topological number1}) can be equivalently expressed as a contour integral of the angular function $\Omega$ introduced in Eq.~(\ref{angle}). To see this, note that the topological current (\ref{jmu}) can be rewritten in terms of a gauge potential $A_\mu=\epsilon_{ab} n^a\partial_\mu n^b$. Using the unit vector $n^a=(\cos\Omega,\sin\Omega)$, one finds $A_\mu=\partial_\mu\Omega$. The topological current density then becomes
\begin{align}
j^0=\frac{1}{2\pi}(\partial_1A_2-\partial_2A_1)=\frac{1}{2\pi}(\partial_1\partial_2\Omega-\partial_2\partial_1\Omega)\;,
\end{align}
which, when integrated over the region $\Sigma$, yields via Stokes’ theorem
\begin{align}
\label{W}
W=\int_{\Sigma}j^0d^2x=\frac{1}{2\pi}\oint_{\partial\Sigma}A_\mu dx^\mu=\frac{1}{2\pi}\oint_{\partial\Sigma} d\Omega\;.
\end{align}
Here $\partial\Sigma=C$ is a counterclockwise closed contour enclosing $\Sigma$. This derivation shows that Eq.~(\ref{W}) computes the same topological invariant as Eq.~(\ref{topological number1}); it provides a practical means of evaluating $W$ by tracking the winding of the vector field $\phi^a$ along the boundary. Below, we choose the contour $C=\sum_i\cup l_i$ as shown in Fig.~\ref{C}. 

To compute $W$, we analyze the behavior of Eqs.~(\ref{phir0}) and (\ref{phit0}) on the boundaries.  At $\theta=0$ and $\pi$,  we have:
\begin{align}
	\phi^r_{l_4}(\theta\to 0^+)\sim \frac{1}{\theta},\quad\phi^\theta_{l_4}(\theta\to 0^+)\sim -\frac{1}{\theta^2}\;,
	\phi^r_{l_2}(\theta\to \pi^-)\sim \frac{1}{\pi-\theta},\quad\phi^\theta_{l_2}(\theta\to \pi^-)\sim \frac{1}{(\pi-\theta)^2}\;,
\end{align}
Thus, the vector $\phi$ points vertically upward at $\theta=0$ and vertically downward at $\theta=\pi$, which is shown in Fig.\ref{C}. The asymptotic behavior of $\phi^r$ and $\phi^\theta$ at $l_1$ and $l_3$ will be studied in the following sections based on our previous work~\cite{Song:2025vhw}.

\section{Between two neighboring horizons}\label{section3}
In this section, we study the topology of circular orbits for spinning test particles between two neighboring horizons.

\begin{figure}[H]
	\centering
	
	\subfigure{
		\begin{minipage}[t]{0.5\linewidth}
			\centering
			\includegraphics[width=2.5in]{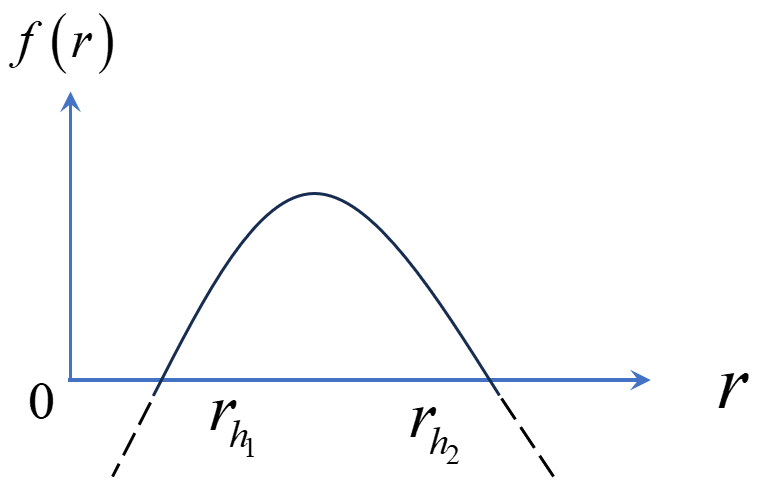}
			\begin{center}
				(a).\, $f'>0$ at $r_{h_1}$, $f'<0$ at $r_{h_2}$ and $f''<0$.
			\end{center}
		\end{minipage}%
	}%
	\subfigure{
		\begin{minipage}[t]{0.5\linewidth}
			\centering
			\includegraphics[width=2.5in]{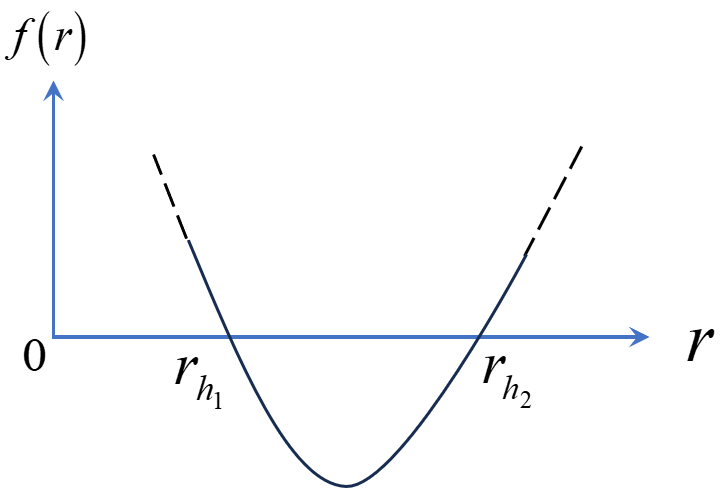}
			\begin{center}
				(b).\, $f'<0$ at $r_{h_1}$, $f'>0$ at $r_{h_2}$ and $f''>0$.
			\end{center}		
		\end{minipage}%
	}%
	\centering
	\caption{The behavior of $f(r)$ in the region between two neighboring horizons.}
	\label{2}
\end{figure}

\begin{figure}[H]
	\centering
	\includegraphics[width=3in]{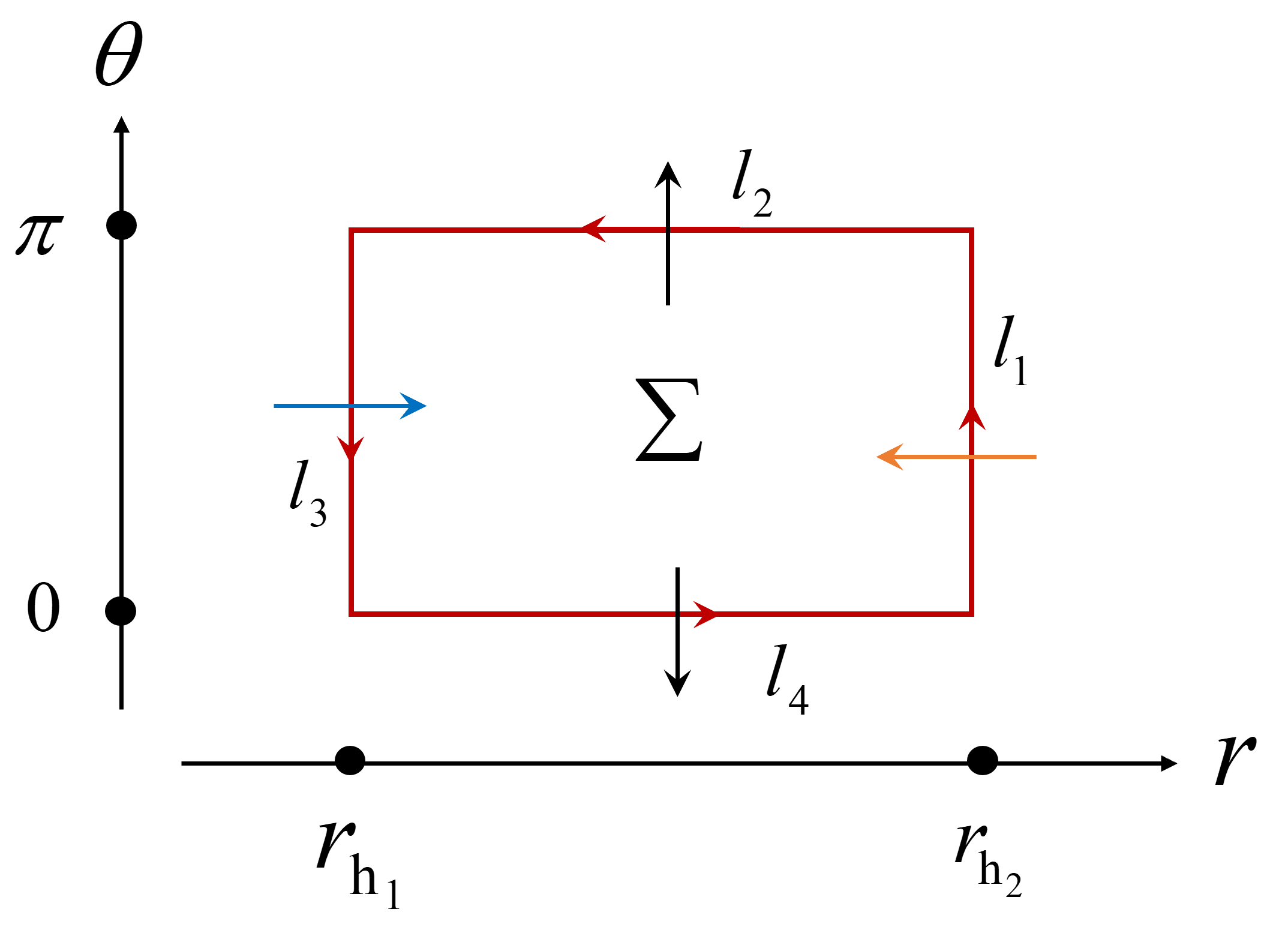}
	\caption{Representation of the contour $C=\sum_il_i$ (which encloses $\Sigma$) on the $(r,\theta)$ plane. The curve $C$ has a positive orientation, marked with the red arrows. The black, blue and yellow arrows indicate the approximate directions of the vector $\phi$ at the boundaries.}
	\label{Ctwohorizons}
\end{figure}

Assume multiple horizons exist, and consider two neighboring horizons $r_{h_1}$ and $r_{h_2}$ with $r_{h_1}<r_{h_2}$. The behavior of $f(r)$ is shown in Fig.~\ref{2}. We consider only case (a), where $f(r)>0$ between the horizons. In case (b), $f(r)<0$ would make $\phi^\theta$ and $\phi^r$ undefined, so circular orbits are forbidden. At the horizon, one has $f(r_h)=0$.
We adopt the contour $C=\sum_{i} l_i$ where $l_1\sim l_4$:$\{r_{\mathrm{out}}=r_{h_2},0\le\theta\le\pi\}\cup\{\theta=\pi, r_{h_1}\le r<r_{h_2}\}\cup\{r_{\mathrm{in}}=r_{h_1},0\le\theta\le\pi\}\cup\{\theta=0,r_{h_1}\le r<r_{h_2}\}$, as  illustrated in Fig.~\ref{Ctwohorizons}.

At $r_{h}$, on the equatorial plane, we have:
\begin{align}
\label{phirrh}
\phi^r_{l_3}(r\to r_{h})\sim\frac{f'(2r-f's^2)\sqrt{L^2+\mathcal{M}^2r^2}}{4r^2}
\end{align}
For the case (a) in Fig.~\ref{2}, at $r_{h_1}$, we have $f'(r_{h_1})>0$. From Eq.~(\ref{conditions}), we find:
\begin{align}
\phi^r_{l_3}(r\to r_{h_1}^+)>0\;,
\end{align}
Ignoring  the specific value of $\phi^\theta$, the vector $\phi$ points to the right on $l_3$ (blue arrow).
	
At $r_{h_2}$, since $f'(r_{h_2})<0$, we have
\begin{align}
\phi^r_{l_1}(r\to r_{h_2}^-)<0\;,
\end{align}
so $\phi$ points to the left on $l_1$ (yellow arrow).
	
Combining the behavior at $\phi$ at $\theta=0$ and $\theta=\pi$, we obtain
\begin{align}
W=\frac{1}{2\pi}\oint d\Omega=\frac{1}{2\pi}\times(-2\pi)=-1\;.
\end{align}
This indicates that for fixed $L$ and $s$, there is always at least one unstable TCO between $r_{h_1}$ and $r_{h_2}$~\cite{Cunha:2020azh,Wei:2022mzv}.  Crucially, this result is independent of the particle's spin magnitude $s$ and orientation (co- or counter-rotating). The topological guarantee of an unstable TCO between horizons has physical consequences for accretion processes and orbital dynamics in multi-horizon spacetimes. The presence of such orbits may influence the stability of matter configurations and energy extraction mechanisms in these regions. Our analysis extends the topological guarantee of unstable TCOs, previously known for non-spinning particles, to the general case of spinning bodies, revealing the geometric origin of this feature.


\section{Outsider the outermost horizon}\label{section4}

\subsection{Asymptotically flat black holes}
\begin{figure}[H]
	\centering
	\includegraphics[width=2.5in]{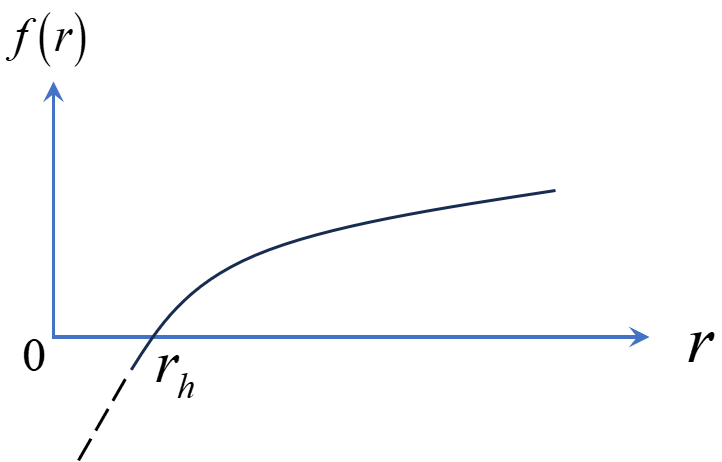}
	\caption{The behavior of $f(r)$  in an asymptotically flat black hole. At $r_h$, one has $f'(r)>0$.}
	\label{fflat}
\end{figure}

\begin{figure}[H]
	\centering
	\includegraphics[width=3in]{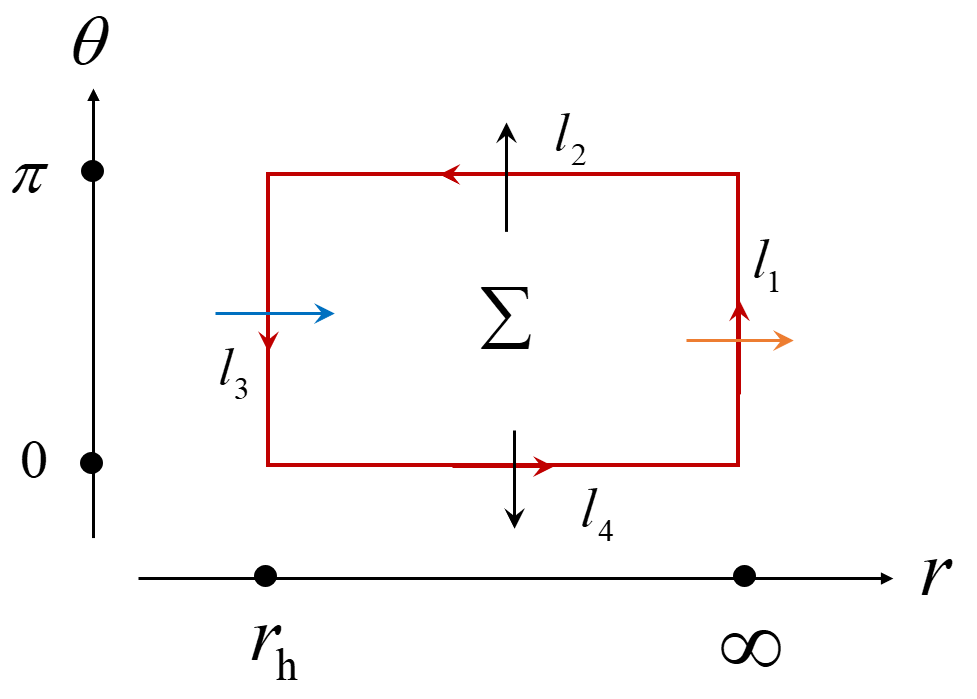}
	\caption{Representation of the contour $C=\sum_il_i$ (which encloses $\Sigma$) on the $(r,\theta)$ plane. The curve $C$ has a positive orientation, marked with the red arrows. The black, blue and yellow arrows indicate the approximate directions of the vector $\phi$ at the boundaries.}
	\label{Cflat}
\end{figure}

In an asymptotically flat black hole described by (\ref{metric}), the metric function behaves as~\cite{Wei:2020rbh}:
\begin{align}
	\label{Aflat}
	f\sim 1-\frac{2M}{r}+\mathcal{O}\bigg(\frac{1}{r^2}\bigg)\;, r\to\infty\;,
\end{align}
where $M$ is the black hole mass. For $r_h<r<\infty$, $f(r)>0$ and $f'(r_h)>0$, as shown in Fig.~(\ref{fflat}).
We choose the contour $C=\sum_{i}l_i$ as $\{r_{\mathrm{out}}=\infty,0\le\theta\le\pi\}\cup\{\theta=\pi, r_h\le r<\infty\}\cup\{r_{\mathrm{in}}=r_h,0\le\theta\le\pi\}\cup\{\theta=0,r_h\le r<\infty\}$, where $r_h$ is the outermost horizon. This is shown in Fig.~\ref{Cflat}.

At $\infty$,  on the equatorial plane:
\begin{align}
\phi_{l_1}^r(r\to\infty)\sim\frac{M\mathcal{M}}{r^2}>0\;,
\end{align}
so $\phi$ points to the right on $l_1$ (yellow arrow).

At $r_{h}$, from Eq.(\ref{phirrh}):
\begin{align}
\phi^r_{l_3}(r\to r_{h}^+)>0\;,
\end{align}
so $\phi$ points to the right on $l_3$ (blue arrow). Combining with the vertical directions at $\theta=0$ and $\theta=\pi$, we find:
\begin{align}
\label{flatW0}
W=\frac{1}{2\pi}\oint d\Omega=\frac{1}{2\pi}\times(\pi-\pi)=0\;.
\end{align}
This implies that, for fixed $L$ and $s$, TCOs must either emerge as a stable–unstable pair or be absent altogether, independent of the particle’s spin.

\begin{itemize}
	\item Example: Schwarzschild black hole
\end{itemize}
For Schwarzschild black hole
\begin{align}
	f(r)=1-\frac{2M}{r}\;,
\end{align}
the effective potential becomes
\begin{align}
\label{Vsch}
V(r)=\frac{Ls(r-3M)r^2+(r^3-Ms^2)\sqrt{(r-2M)[L^2r+\mathcal{M}^2(r^3+2Ms^2-rs^2)]}}{r^2(r^3+2Ms^2-rs^2)}\;.
\end{align}
Set $M = 1$, $\mathcal{M}=0.5$ and $-1\le s\le 1$. The horizon is located at $r_h=2$. The angular momentum $L$ must exceed a critical value to provide sufficient centrifugal support for the existence of stable-unstable TCO pairs, otherwise, no TCOs exist. The critical angular momentum $L_{\mathrm{crit}}$ corresponds to the marginally stable circular orbit (MSCO)~\cite{Wei:2022mzv}. In Schwarzschild black hole, The MSCO is the innermost stable circular orbit (ISCO), and from the condition $V'(r)=V''(r)=0$, one can easily get $L_{\mathrm{ISCO}}\approx 1.732$ for $s=0$. 
\begin{itemize}
	\item [(1).] The presence of $s$ significantly affects the TCOs, such as ISCO, as illustrated in Fig.~\ref{sch_L=1.732}. For $s>0$ (co-rotating), the ISCO bifurcates into a stable–unstable pair of TCOs, whereas for $s<0$ (counter-rotating), the TCOs vanish entirely\footnote{It is important to emphasize that the ISCO itself does not vanish; rather, its position shifts with the spin parameter $s$. Meanwhile, the specific angular momentum originally associated with the ISCO now corresponds either to a stable–unstable TCO pair or to a complete absence of TCOs.}. This indicates that the parameter $s$ modifies the position of the ISCO. Moreover, depending on the magnitude of $s$, this significantly influence can extend to different radial widths regions in the vicinity of the ISCO. For $-1\le s\le 1$, the corresponding range of $L$ is approximately $1.444\lesssim L\lesssim 1.848$. 
	\begin{figure}[H]
		\centering
		\includegraphics[width=7in]{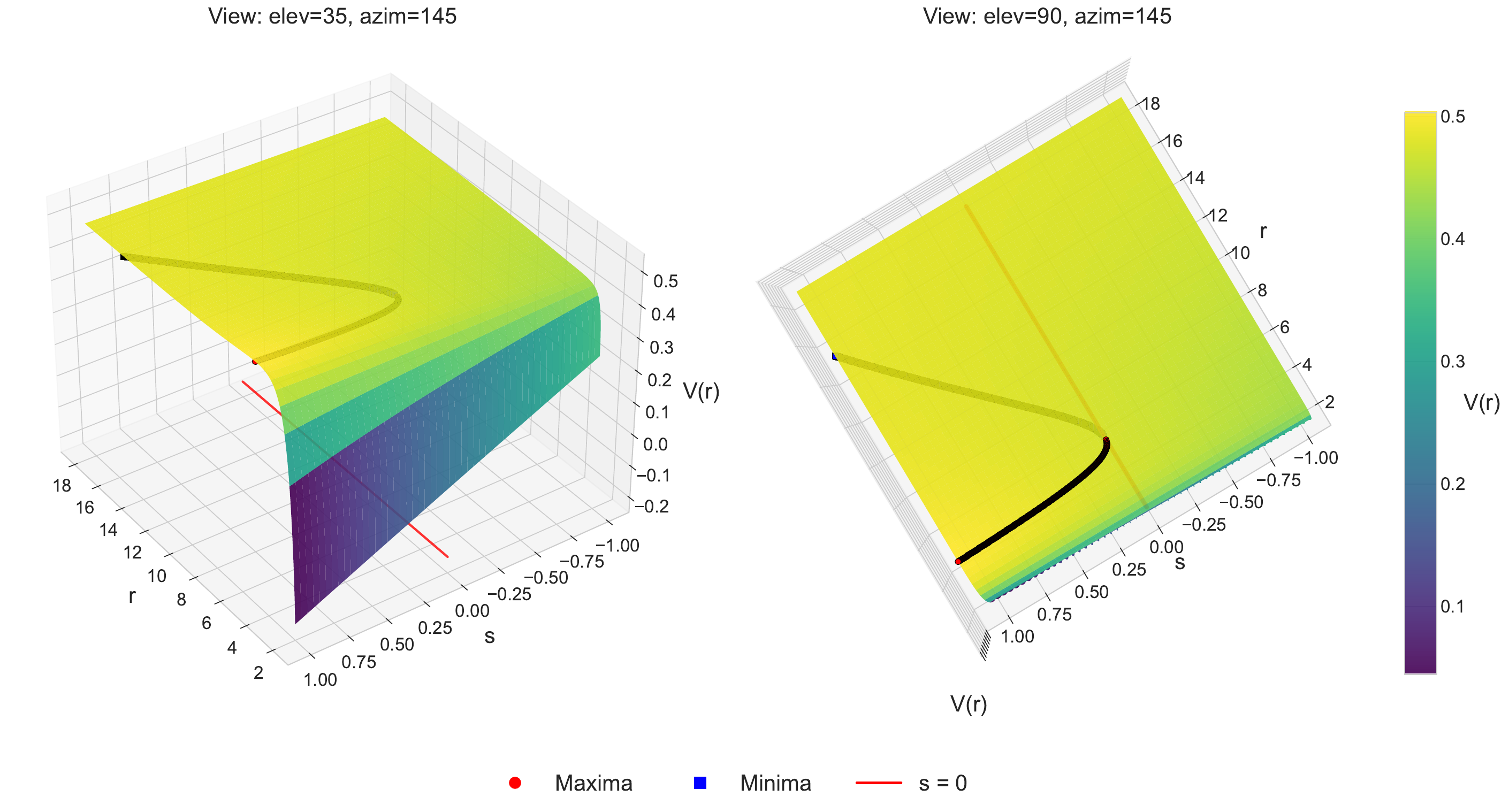}
		\caption{The graph of the potential function (\ref{Vsch}) for $L=1.732$. The red dots and blue squares represent the local maxima and minima of (\ref{Vsch}) for a given value of $s$, respectively. The solid red line corresponds to the locus where $s=0$, which acts as the boundary separating the region without TCOs from the region where such orbits emerge in stable–unstable pairs.}
		\label{sch_L=1.732}
	\end{figure}
\item [(2).] For $L\lesssim 1.444$, no TCOs exist, regardless of the value of $s$ (see Fig.~\ref{sch_L=1}).
\begin{figure}[H]
	\centering
	\includegraphics[width=7in]{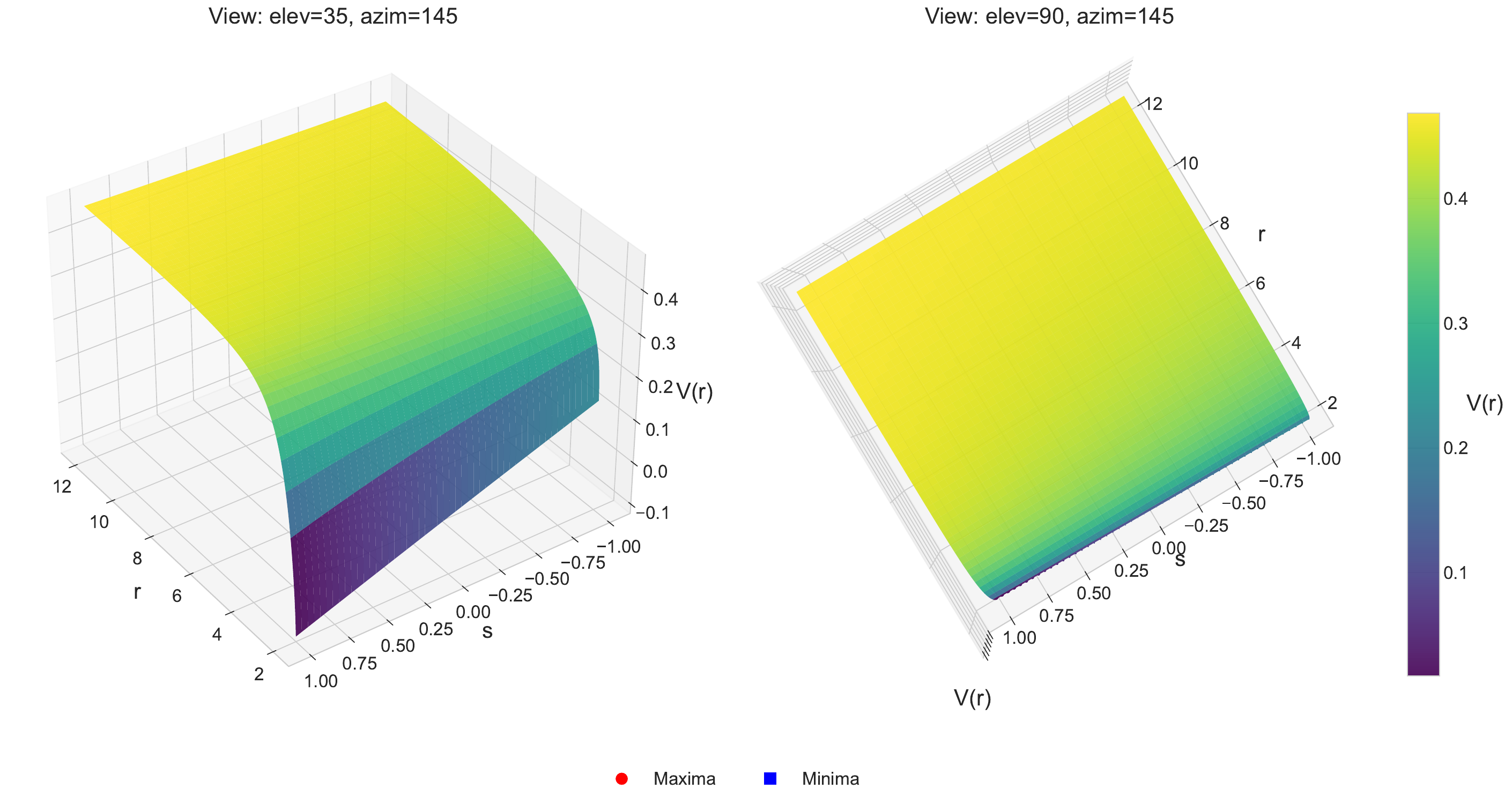}
	\caption{The graph of the potential function (\ref{Vsch}) for $L=1$. The red dots and blue squares represent the maximum values and the minimum values of (\ref{Vsch}) for a given value of $s$, respectively.}
	\label{sch_L=1}
\end{figure}

\item [(3).] For $L\gtrsim 1.848$, pairs of stable and unstable TCOs exist, independent of the value of $s$ (see Fig.~\ref{sch_L=2}).
\begin{figure}[H]
	\centering
	\includegraphics[width=7in]{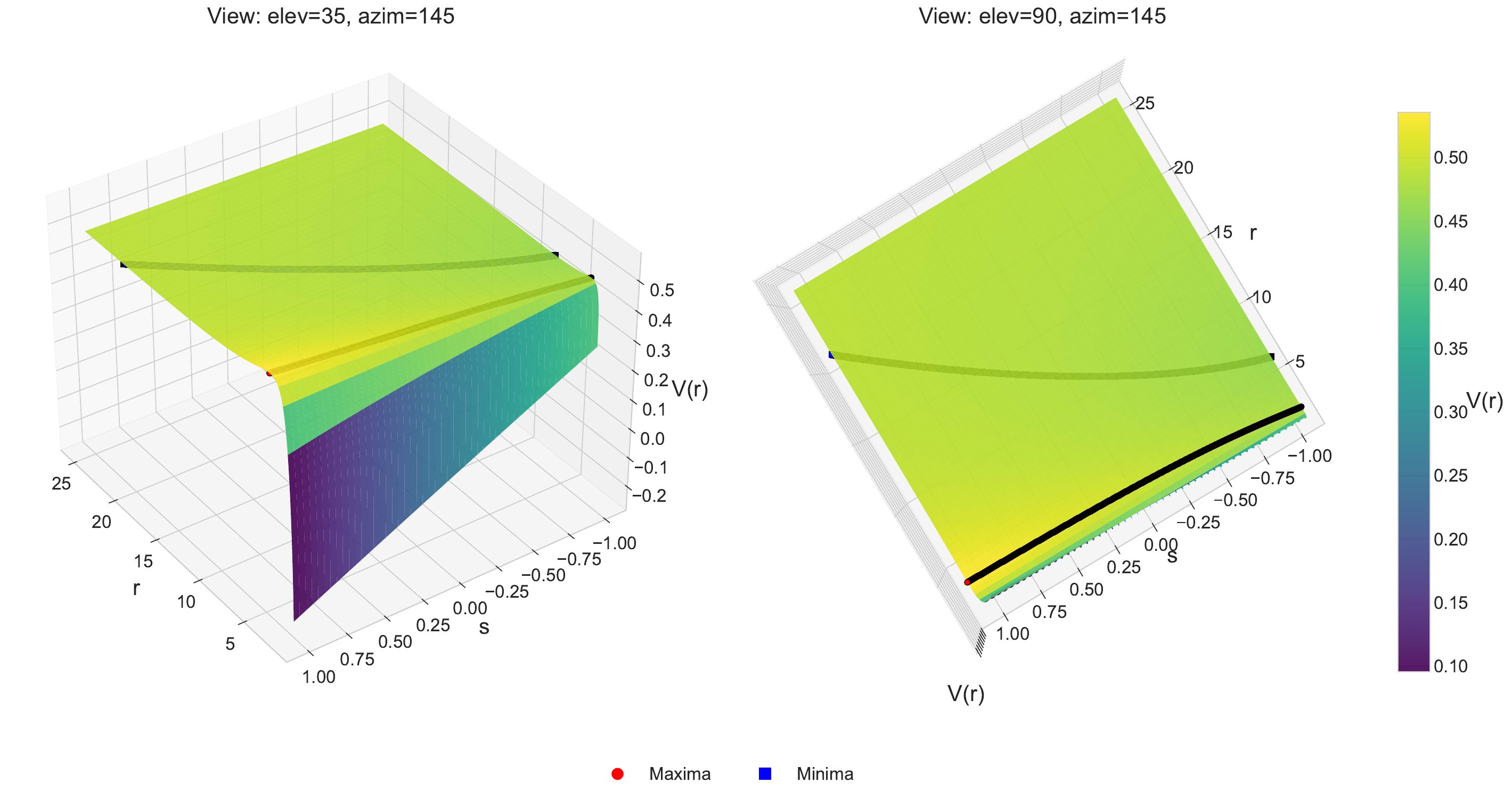}
	\caption{The graph of the potential function (\ref{Vsch}) for $L=2$. The red dots and blue squares represent the maximum values and the minimum values of (\ref{Vsch}) for a given value of $s$, respectively.}
	\label{sch_L=2}
\end{figure}
\end{itemize}
The numerical evidence presented above confirms the topological conclusion that the winding number $W=0$, a result that is independent of the spin parameter. Although the presence of $s$ affects the behavior of TCOs, it does not modify their global topological properties.


\subsection{Asymptotically AdS black holes}
\begin{figure}[H]
	\centering
	\includegraphics[width=2.5in]{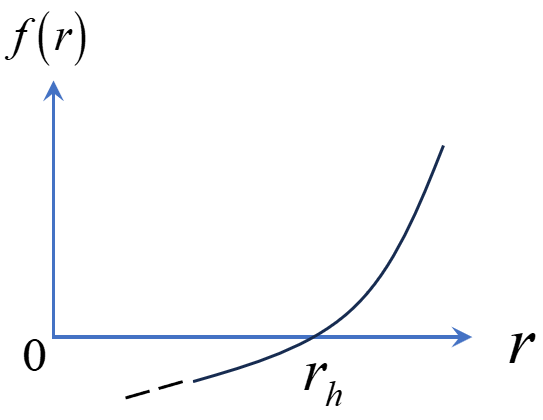}
	\caption{The behavior of $f(r)$  in an asymptotically AdS black hole. At $r_h$, one has $f'(r)>0$.}
	\label{6}
\end{figure}

\begin{figure}[H]
	\centering
	\includegraphics[width=3in]{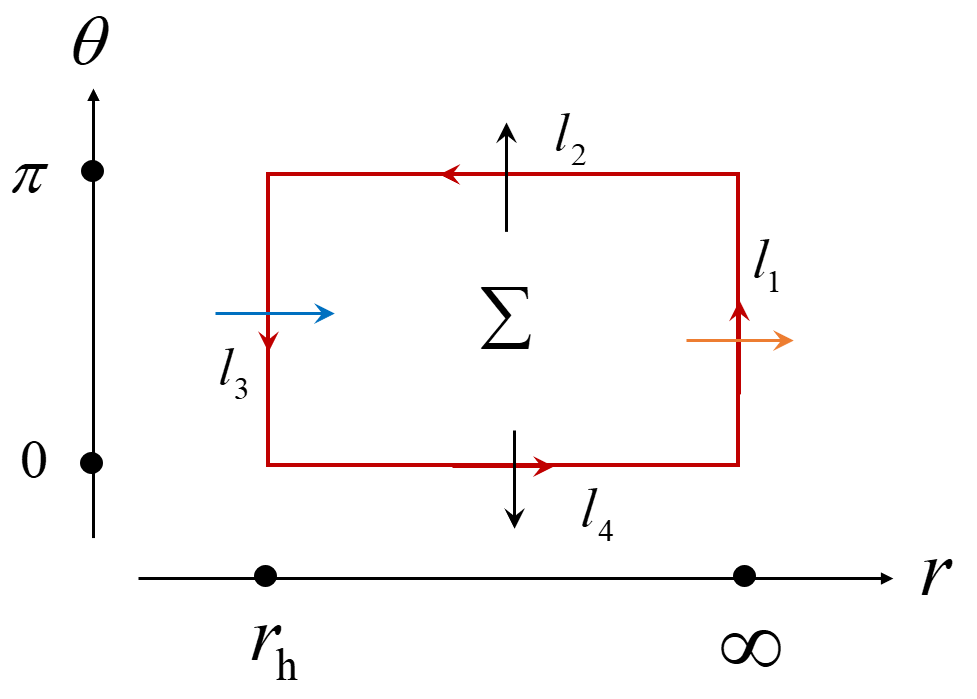}
	\caption{Representation of the contour $C=\sum_il_i$ (which encloses $\Sigma$) on the $(r,\theta)$ plane. The curve $C$ has a positive orientation, marked with the red arrows. The black, blue and yellow arrows indicate the approximate directions of the vector $\phi$ at the boundaries.}
	\label{CAdS}
\end{figure}

For an asymptotically AdS black hole~\cite{Wei:2020rbh}:
\begin{align}
	\label{AAdS}
	&f(r)\sim \frac{r^2}{l^2}+1-\frac{2M}{r}+\mathcal{O}\bigg(\frac{1}{r^2}\bigg)\;,
\end{align}
where $l$ is the AdS radius. For $r_h<r<\infty$, $f(r)>0$ and $f'(r_h)>0$ (Fig.~\ref{6}).
The contour is the same as in the asymptotically flat case (Fig.~\ref{CAdS})

At $\infty$, 
\begin{align}
	\phi_{l_1}^r(r\to\infty)\sim\frac{\mathcal{M}\sqrt{l^2-s^2}}{l^3}r>0\quad (\mathrm{if}\; l>|s|)\;,
\end{align}
so $\phi$ points to the right (yellow arrow).

At $r_{h}$,
\begin{align}
	\phi^r_{l_3}(r\to r_{h}^+)>0\;,
\end{align}
so $\phi$ points to the right (blue arrow).  Thus:
\begin{align}
	\label{AdSW0}
	W=\frac{1}{2\pi}\oint d\Omega=\frac{1}{2\pi}\times(\pi-\pi)=0\;.
\end{align}
This implies that, for fixed $L$ and $s$, TCOs must either emerge as a stable–unstable pair or be absent altogether, irrespective of the particle’s spin, provided $l>|s|$.

\begin{figure}[H]
	\centering
	\includegraphics[width=3.5in]{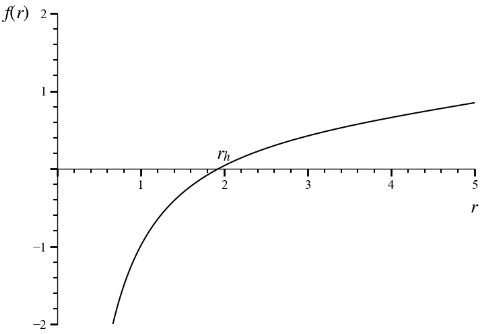}
	\caption{The behavior of $f(r)$ in Sch-AdS black hole. At $r_h$, one has $f'(r)>0$.}
	\label{fAdS}
\end{figure}

\begin{itemize}
	\item Example: Schwarzschild-AdS (Sch-AdS) Black Hole
\end{itemize}
For Sch-AdS:
\begin{align}
	f(r)=1-\frac{2M}{r}+\frac{r^2}{l^2}\;,
\end{align}
Set $M = 1$, $\mathcal{M}=0.5$, $l=10$ and $-1\le s\le 1$. The horizon is located at $r_h\approx 1.9283$. The behavior of $f(r)$ in Sch-AdS black hole is shown in Fig.~\ref{fAdS}.
The effective potential is:
\begin{align}
\label{VAdS}
	V(r)=\frac{(r-3M)l^2Lsr+[l^2r^3-(l^2M+r^3)s^2]\sqrt{(1-\frac{2M}{r}+\frac{r^2}{l^2})\{L^2+\mathcal{M}^2[r^2-(1-\frac{2M}{r}+\frac{r^2}{l^2})s^2]\}}}{l^2r^2[r^2-(1-\frac{2M}{r}+\frac{r^2}{l^2})s^2]}\;.
\end{align}

\begin{itemize}
	\item [(1).] In Sch-AdS black hole, $L_{\mathrm{ISCO}}\approx 2.526$ for $s=0$. The presence of $s$ significantly affects the TCOs, such as ISCO (see Fig.~\ref{sch-AdSL=2.526}). For $-1\le s<0$ (counter-rotating) and $0.5\lesssim s<1$ (co-rotating), the ISCO bifurcates into a stable–unstable pair of TCOs, whereas for $0<s\lesssim 0.5$ (counter-rotating), the TCOs vanish entirely. This indicates that the parameter $s$ modifies the position of the ISCO. Moreover, depending on the magnitude of $s$, this significantly influence can extend to different radial widths regions in the vicinity of the ISCO. For $-1\le s\le 1$, the corresponding range of $L$ is approximately $2.11\lesssim L\lesssim 2.54$.
	\begin{figure}[H]
		\centering
		\includegraphics[width=7in]{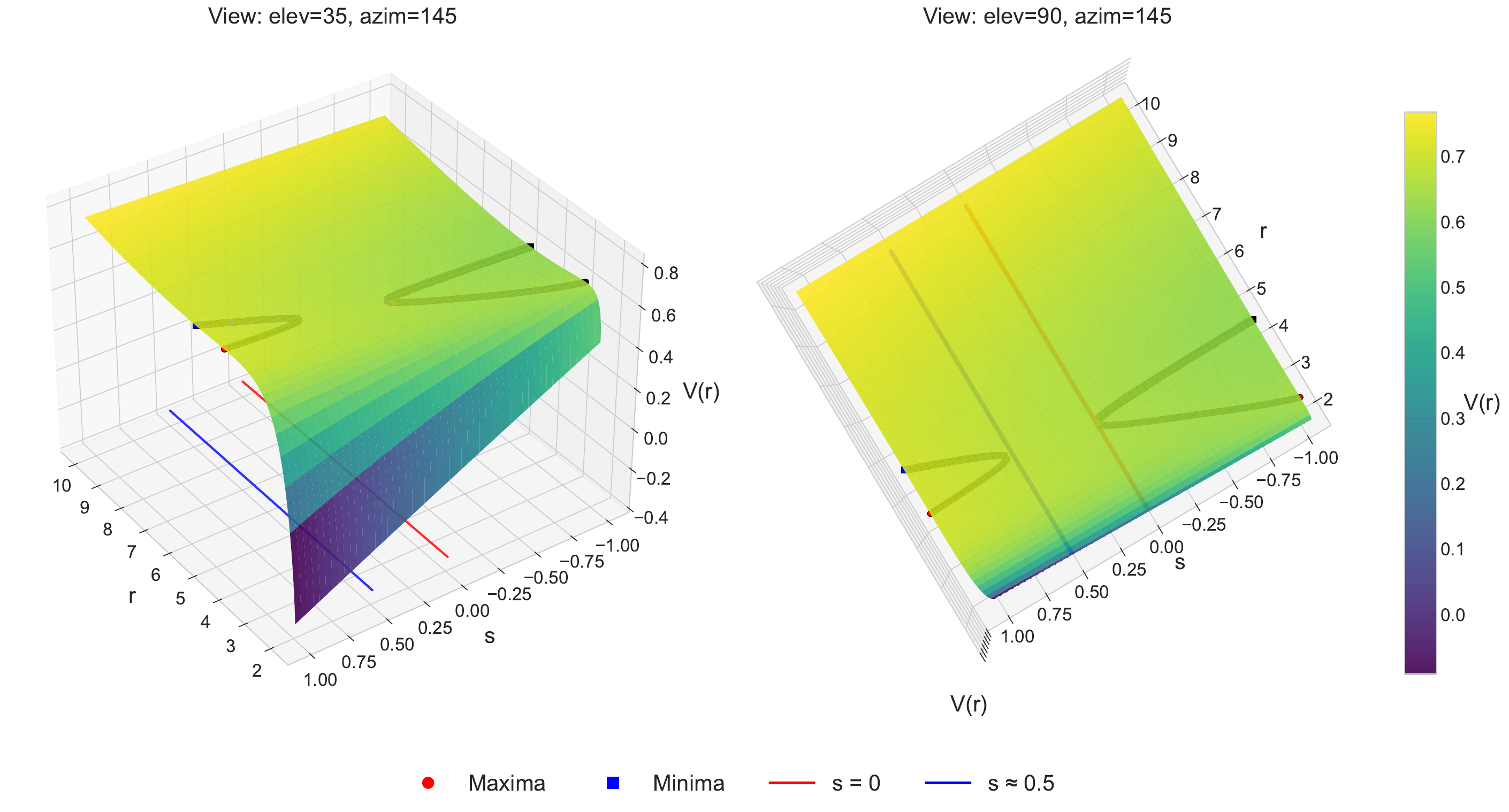}
		\caption{The graph of the potential function (\ref{VAdS}) for $L\approx 2.526$. The red dots and blue squares represent the maximum values and the minimum values of (\ref{VAdS}) for a given value of $s$, respectively. The red and the blue line represent $s=0$ and $s\approx 0.5$, respectively.}
		\label{sch-AdSL=2.526}
	\end{figure}
	
	\item [(2).] For $L\lesssim 2.11$, no TCOs exist, independent of the value of $s$ (see Fig.~\ref{sch-AdSL=1.5}).
	\begin{figure}[H]
		\centering
		\includegraphics[width=7in]{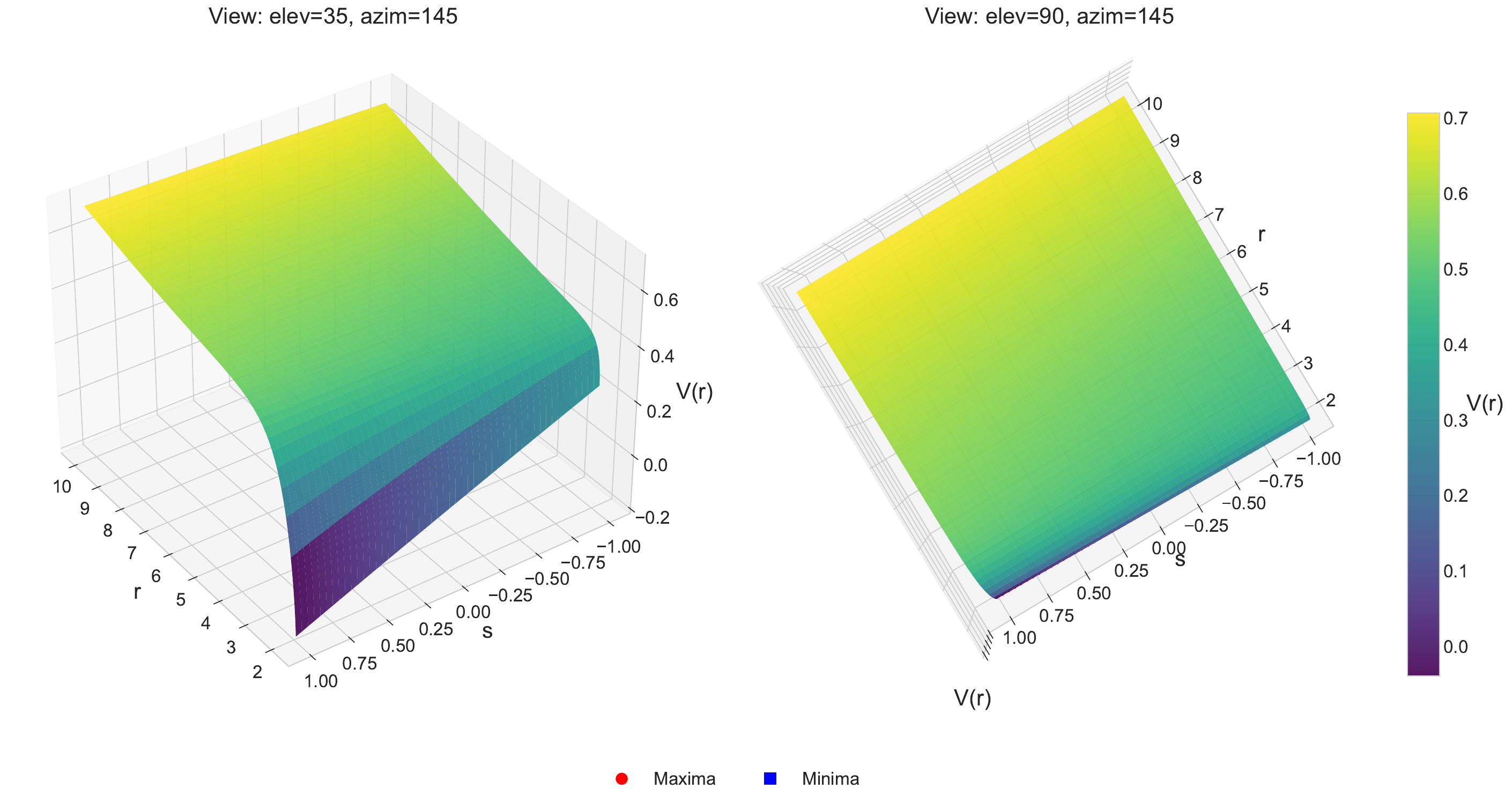}
		\caption{The graph of the potential function (\ref{VAdS}) for $L=1.5$. The red dots and blue squares represent the maximum values and the minimum values of (\ref{VAdS}) for a given value of $s$, respectively.}
		\label{sch-AdSL=1.5}
	\end{figure}
	
	\item [(3).] For $L\gtrsim 3$, pairs of stable and unstable TCOs exist, independent of the value of $s$ (see Fig.~\ref{sch-AdSL=3}).
	\begin{figure}[H]
		\centering
		\includegraphics[width=7in]{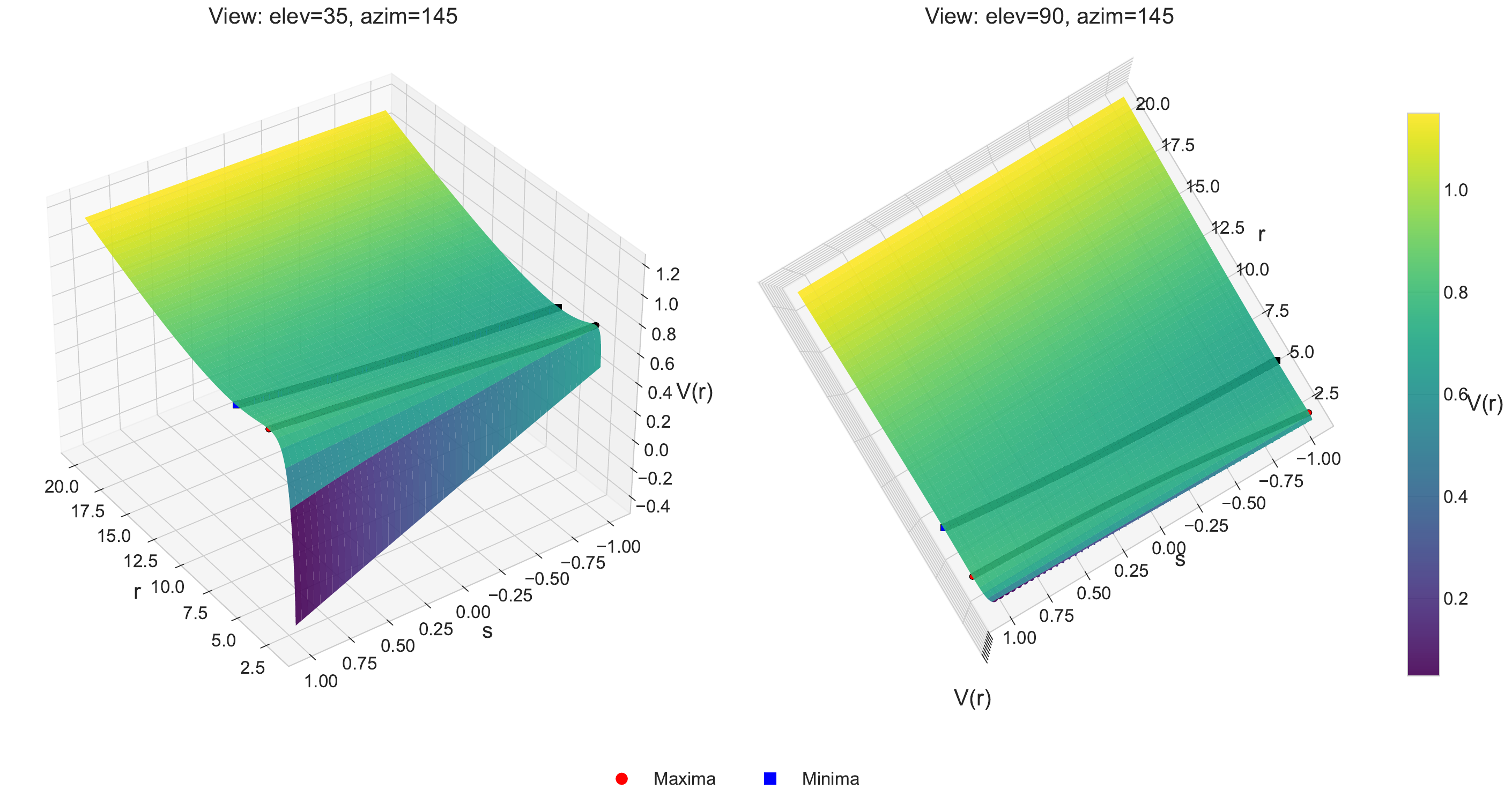}
		\caption{The graph of the potential function (\ref{VAdS}) for $L=3$. The red dots and blue squares represent the maximum values and the minimum values of (\ref{VAdS}) for a given value of $s$, respectively.}
		\label{sch-AdSL=3}
	\end{figure}
\end{itemize}
The numerical evidence presented above confirms the topological conclusion that the winding number $W=0$, a result that is independent of the spin parameter. Although the presence of $s$ affects the behavior of TCOs, it does not modify their global topological properties.

\subsection{Asymptotically dS black holes}
\begin{figure}[H]
	\centering
	\includegraphics[width=2.5in]{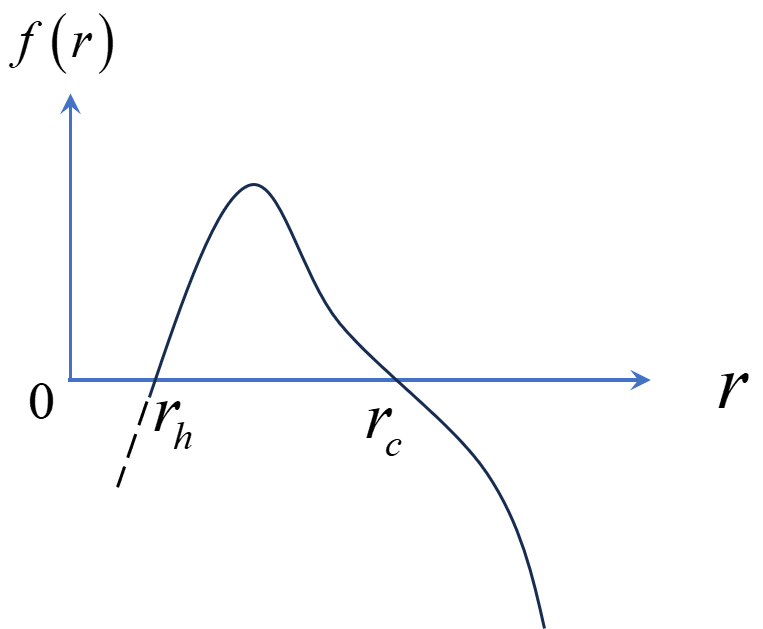}
	\caption{The behavior of $f(r)$ in asymptotically dS black hole. One has $f'(r)>0$ at $r_h$ and $f'(r)<0$ at $r_c$. Here, $r_h$ denotes the outermost black hole horizon, while $r_c$ represents the cosmological horizon.}
	\label{7}
\end{figure}

For an asymptotically dS black hole~\cite{Wei:2020rbh}:
\begin{align}
	f(r)\sim -\frac{r^2}{l^2}+1-\frac{2M}{r}+\mathcal{O}\bigg(\frac{1}{r^2}\bigg)\;,\quad r\to\infty\;,
\end{align}
where $l$ is the curvature radius of dS spacetime. Besides the black hole horizon $r_h$,  there is a cosmological horizon $r_c>r_h$, with $f(r_c)=0$. We consider $r_h<r<r_c$. The behavior of $f(r)$ is as in Fig.~\ref{7} (case (a) of Fig.~\ref{2}). Thus, from Sec.~\ref{section3}, we have:
\begin{align}
\label{dSW-1}
W=-1\;,
\end{align}
indicating at least one unstable TCO for fixed $L$ and $s$. The cosmological horizon in dS spacetime introduces a natural boundary that modifies the topological classification. The guaranteed unstable TCO between the black hole and cosmological horizons may have implications for the global structure of orbits in an expanding universe, particularly for supermassive black holes where this region is astrophysically accessible.

\begin{figure}[H]
	\centering
	\includegraphics[width=3.5in]{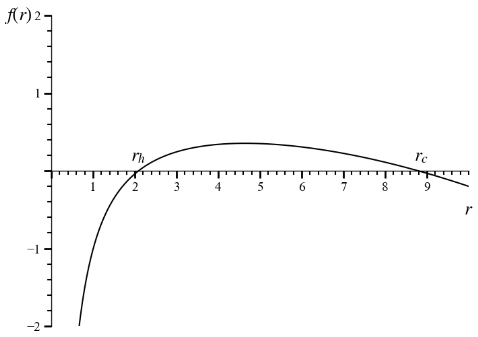}
	\caption{The behavior of $f(r)$ in Sch-AdS black hole. At $r_h$, one has $f'(r)>0$.}
	\label{fdS}
\end{figure}
\begin{figure}[H]
\centering
\includegraphics[width=7in]{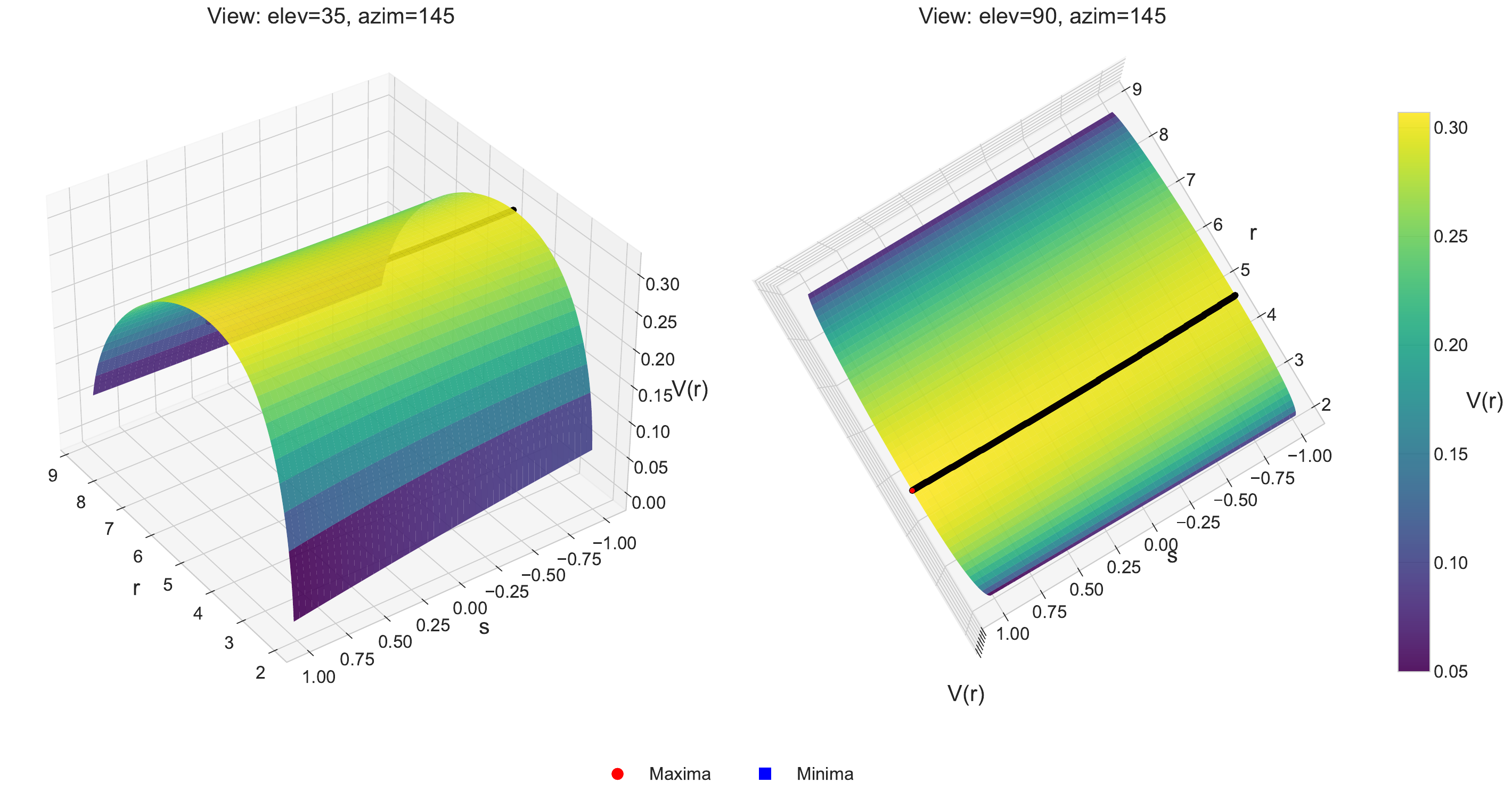}
\caption{The graph of the potential function (\ref{VdS}) for $L=0.3$. The red dots and blue squares represent the maximum values and the minimum values of (\ref{VdS}) for a given value of $s$, respectively.}
\label{sch-dSL=0.3}
\end{figure}

\begin{itemize}
	\item Example: Schwarzschild-dS (Sch-dS) Black Hole
\end{itemize}
For Sch-dS:
\begin{align}
	f(r)=1-\frac{2M}{r}-\frac{r^2}{l^2}\;.
\end{align}
Set $M=1$, $\mathcal{M}=0.5$, $l=10$ and $-1\le s\le 1$. The graph of the function $f(r)$ is shown in Fig.~\ref{fdS}, with $r_h\approx 2.09149$ and $r_c\approx 8.78885$. 
The effective potential is
\begin{align}
\label{VdS}
V(r)=\frac{(r-3M)rl^2Ls+[r^3s^2+l^2(r^3-Ms^2)]\sqrt{(1-\frac{2M}{r}-\frac{r^2}{l^2})[L^2+\mathcal{M}^2(r^2-s^2+\frac{2Ms^2}{r}+\frac{r^2s^2}{l^2})]}}{r^2l^2[r^2-(1-\frac{2M}{r}-\frac{r^2}{l^2})s^2]}\;.
\end{align}
For $L=0.3$, an unstable TCO exists for any value of $s$ (see Fig.~\ref{sch-dSL=0.3}).
The numerical evidence from the graphical results for $L=0.3$ (one unstable TCO) confirms the topological conclusion $W=-1$, which is independent of spin.

\section{Conclusions and discussions}\label{section5}
In this work, we demonstrate the power of topological methods in classifying circular orbits, even for spinning particles. The invariance of $W$ under spin perturbations suggests a deep robustness in the orbital structure dictated by the background metric. This finding has important implications for gravitational wave astronomy, particularly for modeling extreme mass-ratio inspirals where the secondary object may possess significant spin. Our main conclusions are:
\begin{itemize}
	\item [(1).]In regions between two neighboring horizons—such as the inner and outer black-hole horizons or the black-hole and cosmological horizons of de-Sitter spacetime—we obtain $W = -1$, guaranteeing at least one unstable TCO for every fixed pair $(L,s)$. Outside the outermost horizon, by contrast, asymptotically flat and AdS geometries yield $W = 0$, so TCOs must either emerge in stable–unstable pairs or be absent altogether. Numerical tests in Schwarzschild-dS, Schwarzschild, and Schwarzschild-AdS backgrounds corroborate both statements, extending the earlier non-spinning-particle results to the spinning case.
	
	\item [(2).]The topological number $W$ is independent of the spin magnitude $s$ and orientation (co- or counter-rotating). Although the spin parameter $s$ indeed affects the dynamical behavior of TCOs, it does not change their global topological properties.
\end{itemize}

The spin-independence of the topological winding number implies that the fundamental orbital structure--such as the guaranteed existence of unstable TCOs between horizons or the pairing of stable and unstable TCOs outside the outermost horizon--is a property of the spacetime itself, not of the particle’s internal composition. This universality underpins the robustness of gravitational wave signatures from spinning secondaries in EMRIs, suggesting that certain waveform features are more generic than previously assumed.

While our analysis demonstrates the robustness of topological invariants against spin perturbations, several limitations warrant discussion. The Tulczyjew spin-supplementary condition, though widely used, represents one particular choice for closing the MPD equations. Different supplementary conditions may yield quantitatively different orbital parameters, though we expect the topological invariants to remain unchanged due to their fundamental geometric nature.

Additionally, our restriction to static, spherically symmetric spacetimes excludes potentially interesting effects in rotating backgrounds. In Kerr spacetime, the interplay between black hole spin and particle spin may introduce new topological features that merit investigation. The small-spin approximation ($|s|<<M$) employed here ensures the validity of our analysis but leaves open the question of whether similar topological robustness persists for ultra-spinning particles.

Future work should also consider the astrophysical implications of these topological invariants for gravitational wave detection. The guaranteed existence of certain orbital types (e.g., unstable TCOs between horizons) may leave imprints in the gravitational wave signals from extreme mass-ratio inspirals, potentially providing new observational tests of strong-field gravity.

\section*{Acknowledgement}
This work is supported by the Research Start-up Funding of Chengdu University of Technology (Grant No. 10912-KYQD2022-09307).


\end{document}